# Changes in the electronic structure and properties of graphene induced by molecular charge-transfer

**Barun Das[a, b], Rakesh Voggu[a], Chandra Sekhar Rout[a] and C. N. R. Rao[a, b,*]**

[a] Chemistry and Physics of Materials Unit, Jawaharlal Nehru Centre for Advanced Scientific Research, Jakkur P.O., Bangalore -560 064, India.

[b] Solid State and Structural Chemistry Unit, Indian Institute of Science, Bangalore-560012, India.

**Summary**

Interaction with electron -donor and –acceptor molecules such as aniline and nitrobenzene brings about marked changes in the Raman spectrum and the electronic structure of graphene, prepared by the exfoliation of graphitic oxide.

---

* For correspondence: cnrrao@jncasr.ac.in, Fax: (+91)80-22082766

Graphene is a fascinating two-dimensional nanomaterial with unique electronic structure and properties.[1-3] The Fermi energy ($E_F$) in single-layer graphene is proportional to the square root of the carrier concentration in the plane of the sheet. The Fermi energy is shifted by doping due to stiffening or softening of phonons and other effects which modify the phonon dispersion by changing the carrier concentration and mobility.[4-6] Significant changes in the properties of graphene, in particular its phonon spectrum and electronic structure, are reported to occur when electrons or holes are introduced by electrochemical means.[6] Investigations of single-walled nanotubes (SWNTs) have revealed significant changes in the electronic structure and properties depending on the geometry, doping and chemical environment.[7-12] It has been shown recently that molecules which act as electron-donors or –acceptors modify the electronic structure of SWNTs, giving rise to significant changes in the electronic and Raman spectra as well as electrical properties.[11, 12] Prompted by the results obtained with SWNTs, we have investigated the effect of interaction of electron-donor and –acceptor molecules on the electronic structure and properties of graphene. For this purpose, we have prepared graphene samples by the exfoliation of the graphitic oxide[13, 14] and studied the interaction of monosubstituted benzenes such as nitrobenzene, chlorobenzene, anisole and aniline by employing Raman spectroscopy and electrical resistivity measurements. Raman spectroscopy provides the most useful signature for examining the changes brought about in the electronic structure of graphene. In particular, the G- (1573 $cm^{-1}$) and 2D- (2650 $cm^{-1}$) bands are useful in understanding the effects of electron- and hole-doping. Note that the 2D-band is prominent in the Raman spectrum of graphene while the D-band is absent in the single-layer material.

We have prepared graphene by the exfoliation of graphite oxide by employing the literature procedure [13, 14] and have characterized these samples by powder X-ray diffraction, Raman spectroscopy and atomic force microscopy.[†] The number of layers in the graphene samples prepared by us was around four. In Fig. 1(a), we show the G-band of the graphene after interaction with 1 M solutions of various monosubstituted benzenes with electron-donating and -withdrawing groups.[†] While nitrobenzene causes stiffening of the G-band or an increase in the frequency, aniline causes softening or a shift to a lower frequency. The stiffening or softening of the G-band depends on the electron-donating or –withdrawing power of the substituent on benzene. In Fig. 2(a) we have plotted the position of the G-band maximum against the Hammett σ substituent constant to show how the frequency decreases with the decreasing electron–withdrawing power or increasing electron-donating power of the substituent. Encouraged by this result, we have examined the dependence of the G-band on the concentrations of nitrobenzene and aniline in benzene solutions. Fig. 1(b) shows the G-bands at different concentrations of aniline and nitrobenzene. We show the variation in the position of the G-band maximum with the concentration of aniline and nitrobenzene in Fig 3(a). Interaction with aniline and nitrobenzene causes shifts in the opposite directions, the magnitude of the shift increasing with the concentration. Just as in the case of SWNTs[15], we can consider the interaction with nitrobenzene to be equivalent to hole-doping and that with aniline to equivalent to electron-doping. The 2D-band position also varies on interaction with aniline and nitrobenzene, the latter causing a marked increase in the frequency. Interestingly, the full-width at half maximum (FWHM) of the G-band increases with both

electron- and hole- doping or with the concentration of aniline and nitrobenzene, as can be seen from the inset in Fig 3(a).

The intensity of the 2D-band also changes with the concentration of aniline and nitrobenzene. Thus, we do not see the 2D-band in the Raman spectrum of graphene on interaction with pure aniline or nitrobenzene. The 2D-band appears only when the concentration of these compounds is decreased as shown in the inset of Fig 3(b). The ratio of the intensities of 2D- and G- bands (I(2D)/I(G)) is considered to be sensitive to doping.[6] Fig. 3(b) shows the plot of I(2D)/I(G) against the concentration of aniline and nitrobenzene. The intensity ratio shows a marked decrease with the increase in concentration of electrons as well as of holes, reducing to zero at high concentrations of aniline and nitrobenzene. The I(2D)/I(G) ratio obtained with 1M solutions of monosubstituted benzene shows a maximum when plotted against the Hammett σ substituent constant as shown in Fig. 2(b), indicating that both electron-donating and electron-withdrawing substituents cause a decrease in the intensity ratio. The present study clearly establishes that the Raman spectrum of graphene is sensitive to molecular charge-transfer, the effects being comparable to those obtained by electrochemical doping.[6]

In Fig. 4(a) we show the I-V characteristics of the graphene films after treatment with 1M solutions of monosubstituted benzenes. The I-V characteristics remain linear showing metallic behavior. The resistance itself is lowest in the presence of nitrobenzene and highest in the presence of aniline. There is a systematic dependence of resistance with the electron-donating and –withdrawing power of the substituents. The value of resistance varies with the concentration of aniline and nitrobenzene as shown in Fig. 4(b).

At a bias voltage of 0.5 V, the resistance of the graphene is ~1.0 kΩ. The resistance increases linearly with increasing aniline concentration, while it decreases abruptly at low concentrations of nitrobenzene. Thus, hole-doping brought about by interaction with nitrobenzene has a marked effect even at low concentrations.

In conclusion, the present study demonstrates how the electronic structure and phonons of graphene are markedly affected by interaction with electron-donor and –acceptor molecules. It is significant that we observe such marked effects due to molecular charge-transfer even with multi-layered graphene (average 4 layers). These effects would be expected to be prominent in single-layered graphene as well. Comparing our results with those reported for electrochemically doped single-layer graphene [6], it appears that only static contributions are involved in the spectral changes observed by us. Dynamic contributions may become negligible due to defects in multilayered graphene.

## Notes and references

† Graphite oxide (GO) was synthesized by employing the literature procedure [13, 14]. Briefly, a reaction flask containing a magnetic stir bar was charged with sulfuric acid (18 mL) and fuming nitric acid (9 mL) and cooled by immersion in an ice bath. The acid mixture was stirred and allowed to cool for 20 min, and graphite microcrystals (0.5 g) was added under vigorous stirring to avoid agglomeration. After the graphite powder was well dispersed, potassium chlorate (10 g) was added slowly over 5 min to avoid sudden increases in temperature. The reaction flask was loosely capped to allow evolution of gas from the reaction mixture and allowed to stir for 120 h at room temperature. The resulting product was suction filtered and washed thoroughly with distilled water. The product was dried under vacuum for 24 hours. The graphite oxide so obtained was exfoliated in a furnace preheated to 1050$^{o}$C under argon flow for about 30s.

The graphene samples were characterized using transmission electron microscopy, atomic force microscopy and powder x-ray diffraction. Raman spectra were recorded with a LabRAM HR high-resolution Raman spectrometer (Horiba-Jobin Yvon) using a He−Ne laser ($\lambda$ = 632.8 nm). For Raman measurements, one milligram of the graphene sample was dispersed in 3 ml of benzene containing appropriate concentrations of the monosubstituted benzene and sonicated. The resulting solution was filtered through an anodisc filter (Anodisc 47, Whatman) with a pore size of 0.1 μm. We have carried out electrical resistivity measurements by drop-coating the graphene sample on Au-gap electrodes patterned on glass substrates.

## Figure captions:

Fig. 1: Raman shift of the G-band of graphene on interaction with (a) 1M solutions of monosubstituted benzenes and (b) with varying concentrations aniline and nitrobenzene

Fig. 2: Variation of (a) the G-band frequency and (b) the ratios of the 2D/G bands against the Hammett substituent constant, σ.

Fig. 3: Variation of (a) the G-band Raman shift and (b) the intensity ratio of the 2D/G bands with the concentration of aniline and nitrobenzene. Inset in Fig. 3(a) shows the variation of the FWHM of the G-band against the concentration of aniline and nitrobenzene. Inset in (b) shows 2D bands at different concentrations of aniline and nitrobenzene

Fig. 4: (a) I-V characteristics of the graphene in the presence of benzene and 1 M solutions of nitrobenzene, chlorobenzene, anisole and aniline in benzene (b) Variation of the resistance with the concentration of nitrobenzene and aniline at a bias voltage of 0.5 V.

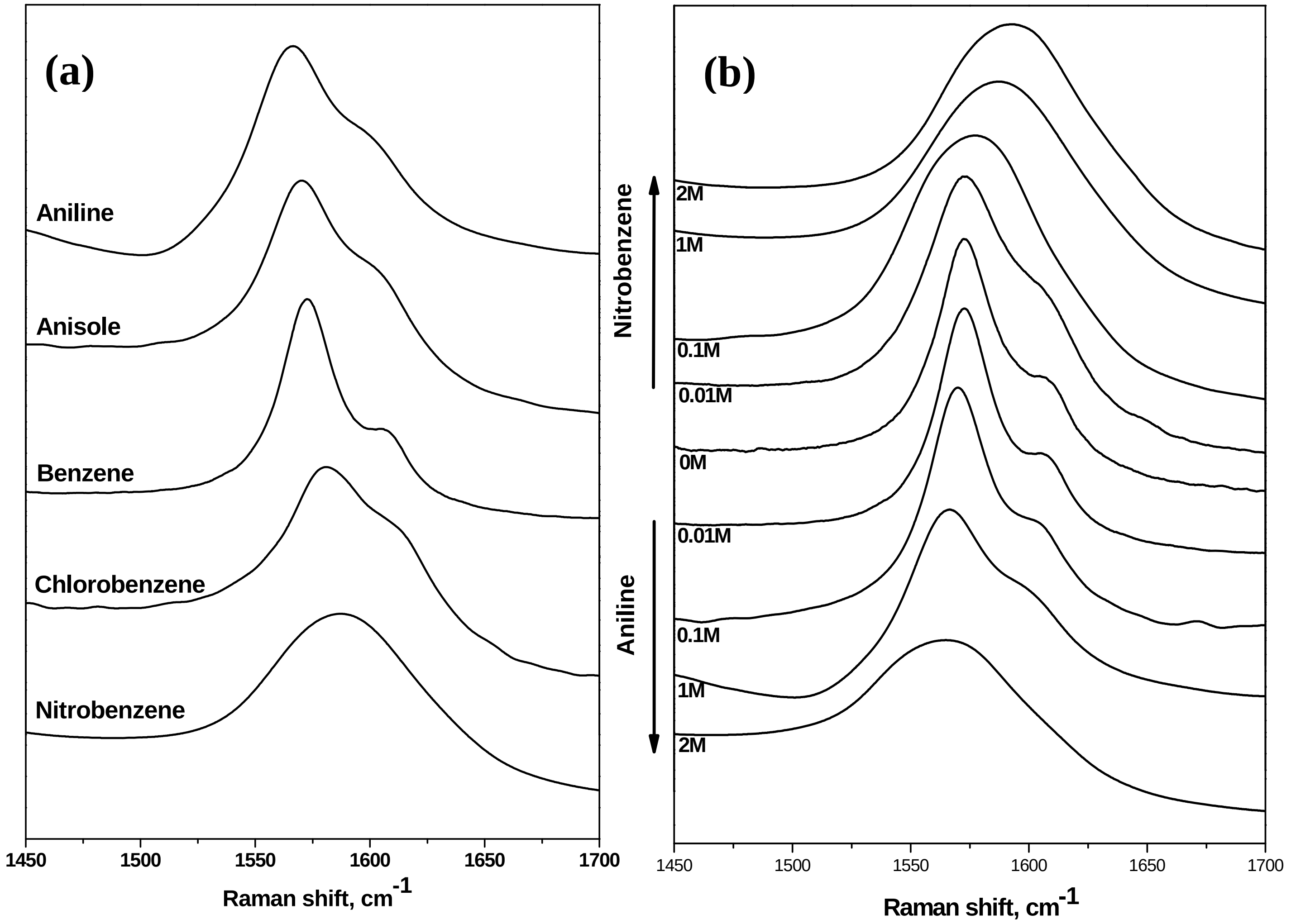

**Figure 1**

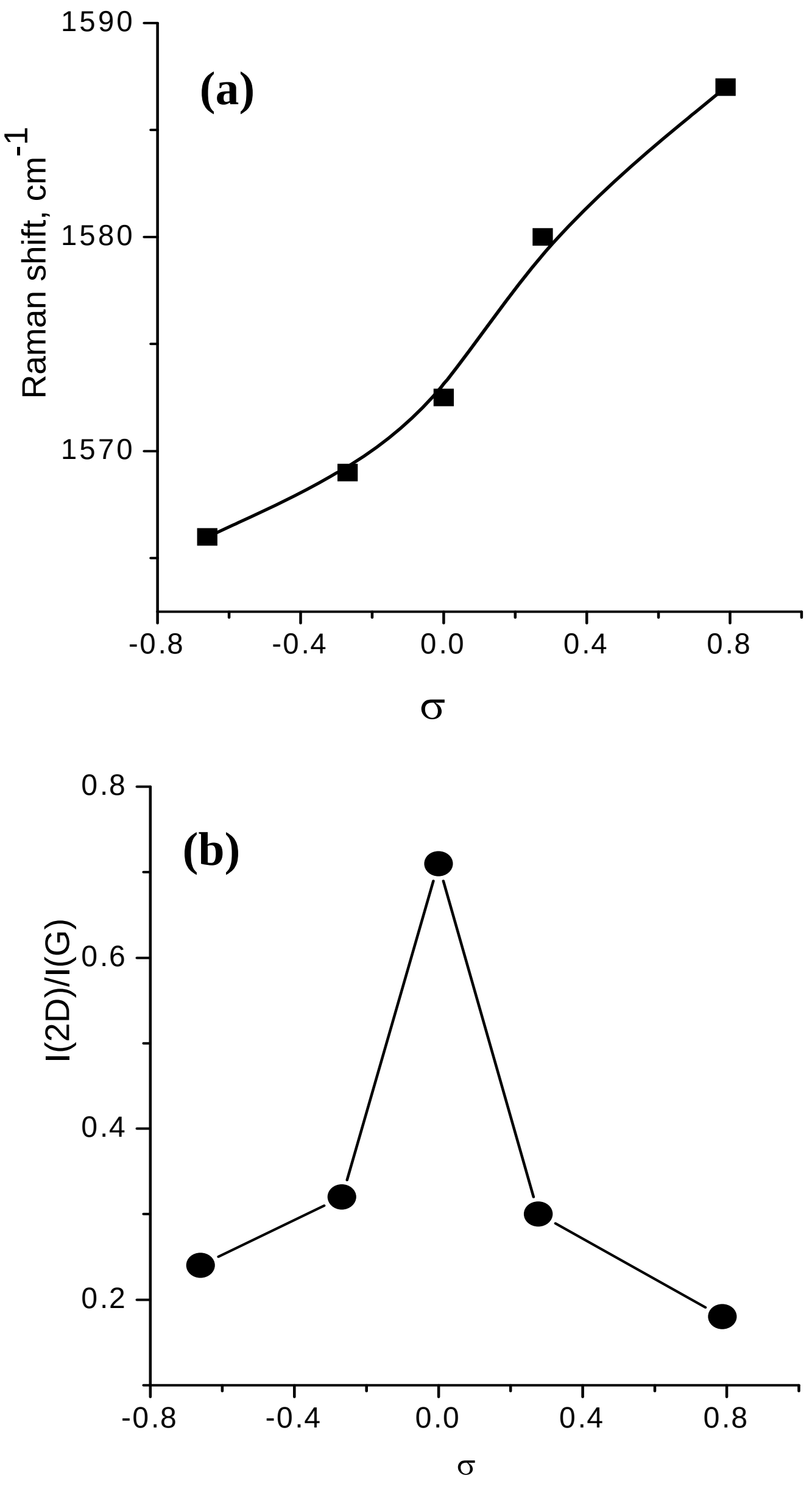

**Figure 2**

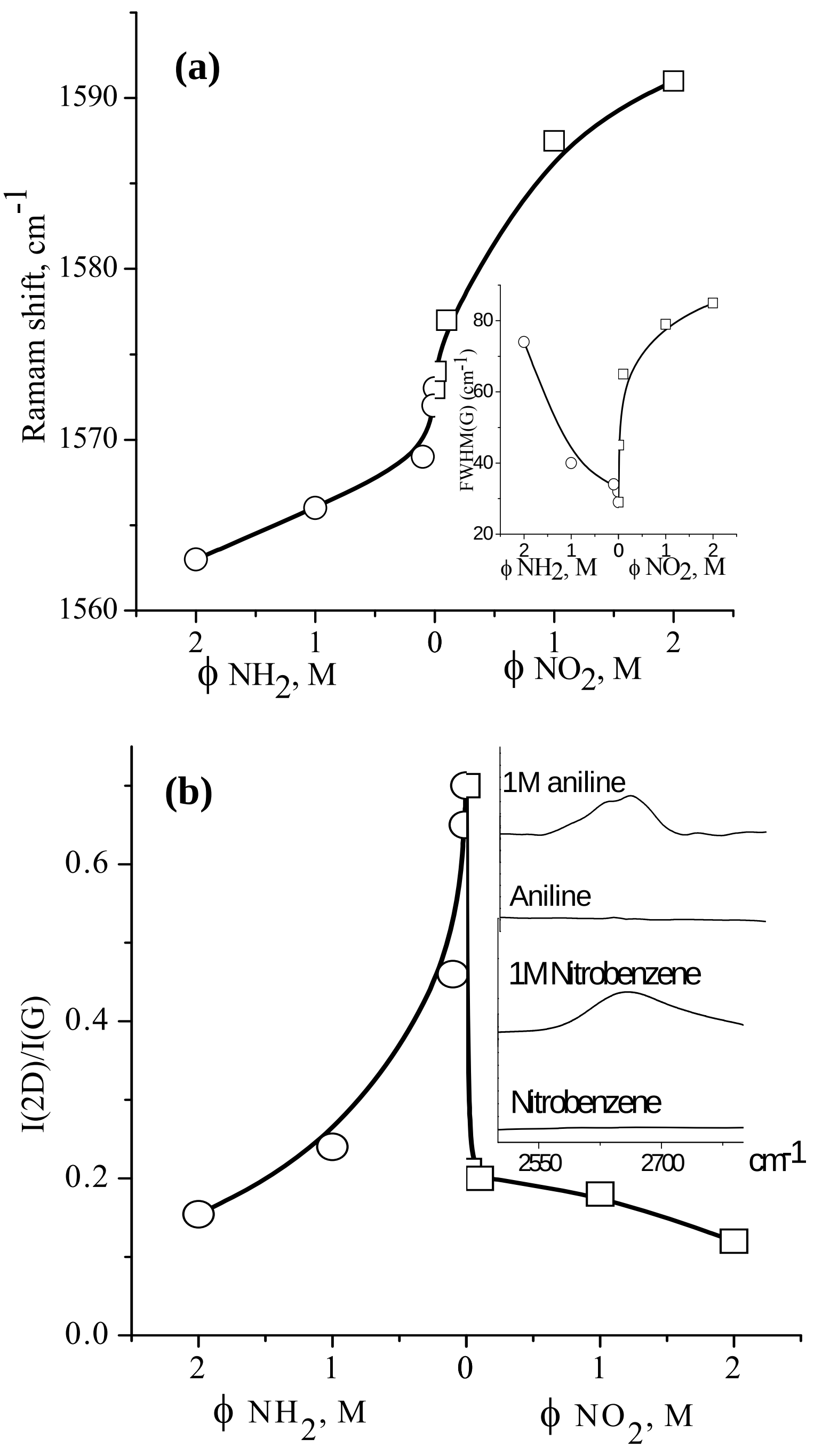

**Figure 3**

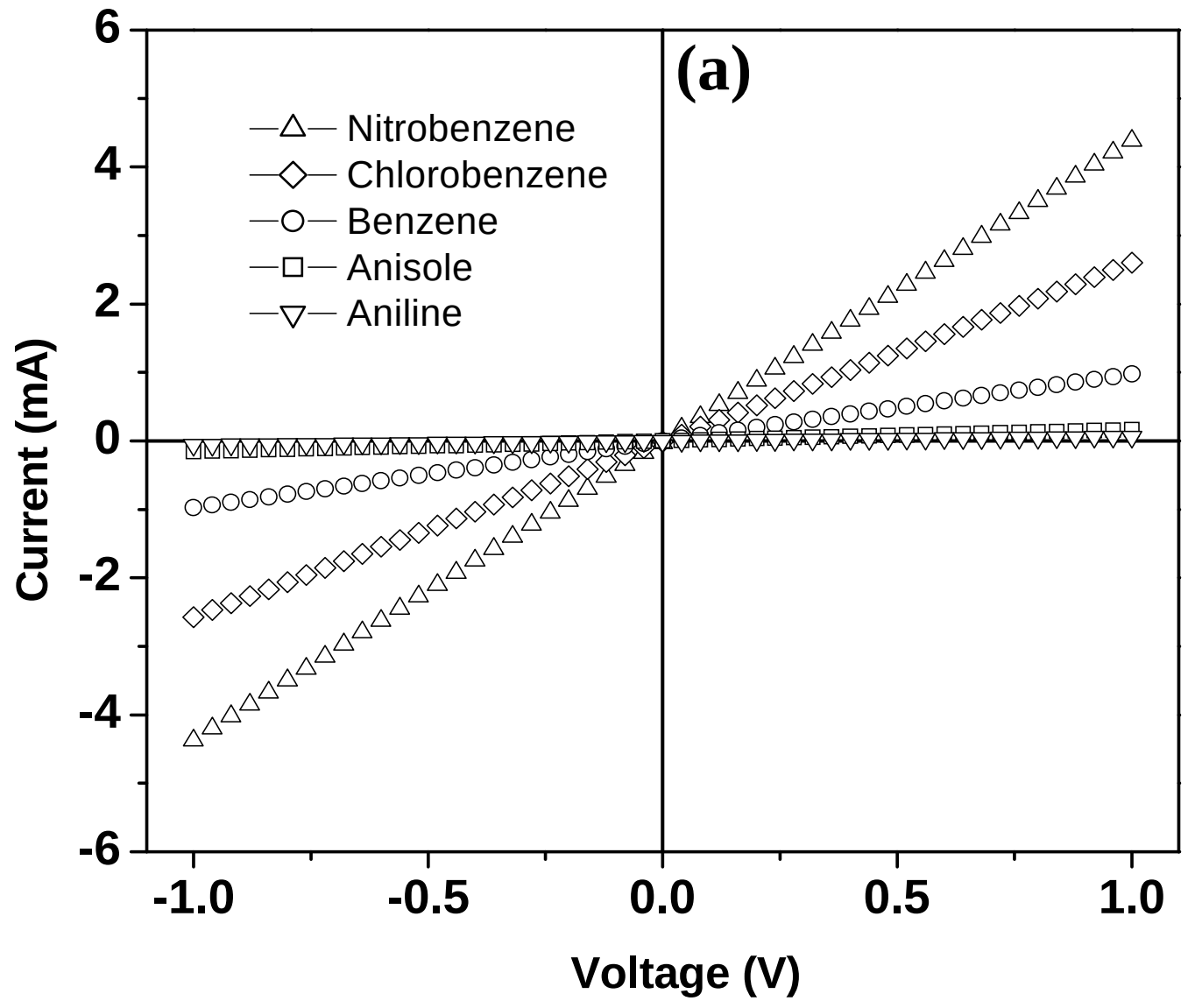

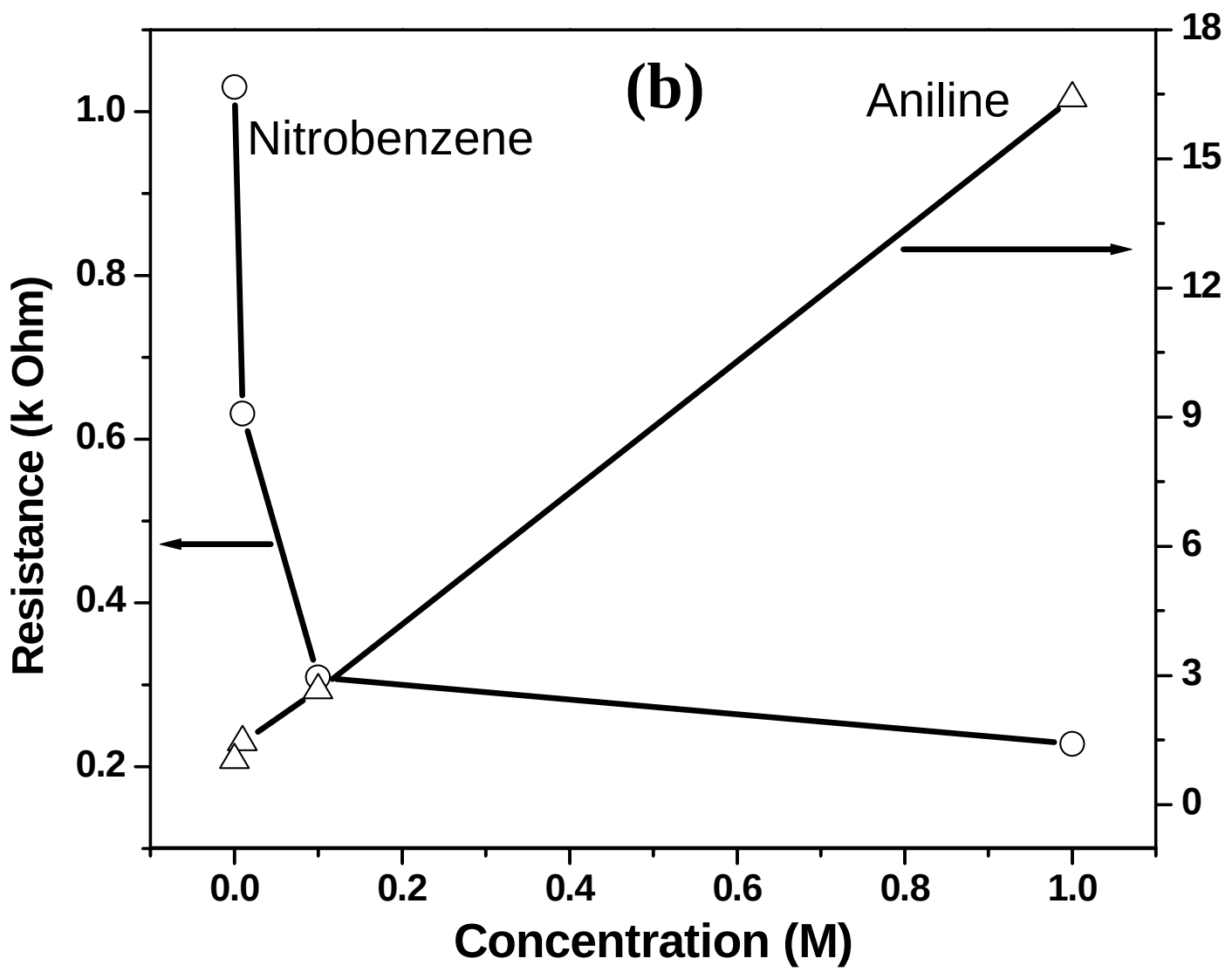

# Figure 4